# Dynamic stabilization of plasma instability

S. Kawata[1], T. Karino[1], and Y. J. Gu[2]

[1] Graduate School of Engineering, Utsunomiya University, 321-8585 Utsunomiya, Japan

[2] ELI-Beamlines, Institute of Physics, 182 21 Prague, Czech Republic

Corresponding author:
Shigeo Kawata, Prof. Dr.
Graduate School of Engineering,
Utsunomiya University
Yohtoh 7-1-2, Utsunomiya 321-8585, Japan.
TEL & FAX: +81-28-689-6080
E-MAIL: kwt@cc.utsunomiya-u.ac.jp

**Abstract**

The paper presents a review of dynamic stabilization mechanisms for plasma instabilities. One of the dynamic stabilization mechanisms for plasma instability was proposed in the papers [Phys. Plasmas 19, 024503(2012) and references therein], based on a perturbation phase control. In general, instabilities emerge from the perturbations of the physical quantity. Normally the perturbation phase is unknown so that the instability growth rate is discussed. However, if the perturbation phase is known, the instability growth can be controlled by a superimposition of perturbations imposed actively: if the perturbation is introduced by, for example, a driving beam axis oscillation or so, the perturbation phase can be controlled and the instability growth is mitigated by the superimposition of the growing perturbations. Based on this mechanism we present the application results of the dynamic stabilization mechanism to the Rayleigh-Taylor (R-T) instability and to the filamentation instability as typical examples in this paper. On the other hand, in the paper [Comments Plasma Phys. Controlled Fusion 3, 1(1977)] another mechanism was proposed to stabilize the R-T instability based on the strong oscillation of acceleration, which was realized by the laser intensity modulation in laser inertial fusion [Phys. Rev. Lett. 71, 3131(1993)]. In the latter mechanism, the total acceleration strongly oscillates, so that the additional oscillating force is added to create a new stable window in the system. Originally the latter mechanism was proposed by P. L. Kapitza, and it was applied to the stabilization of an inverted pendulum. In this paper we review the two dynamic stabilization mechanisms, and present the application results of the former dynamic stabilization mechanism.

# 1. Introduction

Dynamic stabilization mechanisms for plasma instabilities are reviewed and discussed in this paper. So far, the dynamic stabilization for the Rayleigh-Taylor instability (RTI) [1-6] has been proposed and discussed intensively in order to obtain a uniform compression [7, 8] of a fusion fuel pellet in inertial confinement fusion. The RTI dynamic stabilization was found many years ago [1, 2] and is important in inertial fusion. It was implemented that the oscillation amplitude of the driving acceleration should be sufficiently large to stabilize RTI [1-6]. In inertial fusion, the fusion fuel compression is essentially important to reduce an input driver energy [7, 8], and the implosion uniformity is one of critical issues to compress the fusion fuel pellet stably [9, 10]. Therefore, the RTI stabilization or mitigation is attractive to minimize the fusion fuel mix.

On the other hand, instability grows from a perturbation in general, and normally the perturbation phase is unknown. Therefore, it would be difficult to control the perturbation phase, and usually the instability growth rate is discussed. However, if the perturbation phase is controlled and known, we can find a new way to control the instability growth. One of the most typical and well-known mechanisms is the feedback control in which the perturbation phase is detected and the perturbation growth is controlled or mitigated or stabilized. In plasmas it is difficult to detect the perturbation phase and amplitude. However, even in plasmas, if we can actively impose the perturbation phase by the driving energy source wobbling or so, and therefore, if we know the phase of the perturbations, the perturbation growth can be controlled in a similar way as shown in Fig. 1 [11, 12]. In instabilities, one mode of an initial perturbation, from which an instability grows, may have the form of $a = a_0 e^{ikx+\gamma t}$, where $a_0$ is the amplitude, $k = 2\pi/\lambda$ is the wave number, $\lambda$ the wave length and $\gamma$ the growth rate of the instability. An example initial perturbation is shown in Fig. 1(a). At $t=0$ the perturbation is imposed. The initial perturbation may grow at instability onset. After $\Delta t$, if the feedback control works on the system, another perturbation, which has an inverse phase with the detected amplitude at $t=0$, is actively imposed (see Fig. 1(b)), so that the actual perturbation amplitude is very well mitigated as shown in Fig. 1(c). This is an ideal example for the instability mitigation. This control mechanism is apparently different from the dynamic stabilization mechanism shown in the previous works in Refs. [1-6]. For example, the growth of the filamentation instability [13-17] driven by a particle beam or jet could be controlled by the beam axis oscillation or wobbling. The oscillating and modulated beam induces the initial perturbation and also could define the perturbation phase. Therefore, the successive phase-defined perturbations are superimposed, and we can use this property to mitigate the instability growth. Another example can be found in heavy ion beam inertial fusion; the heavy ion accelerator could have a capability to provide a beam axis wobbling with a high frequency [18-20]. The wobbling heavy ion beams also define the perturbation phase. This means that the perturbation phase is known, and so the successively imposed perturbations are superimposed on plasmas. We can again use the capability to reduce the instability growth by the phase-controlled superposition of perturbations. In this paper we discuss and clarify the dynamic mitigation mechanisms for plasma instabilities. First, we discuss the dynamic stabilization mechanism based on Refs. [1-6, 23] to stabilize the RTI by applying the strong and rapid acceleration oscillation. Then we present the other dynamic stabilization mechanism proposed in Refs. [12, 20-22], which is applied to the RTI and filamentation instabilities stabilization.

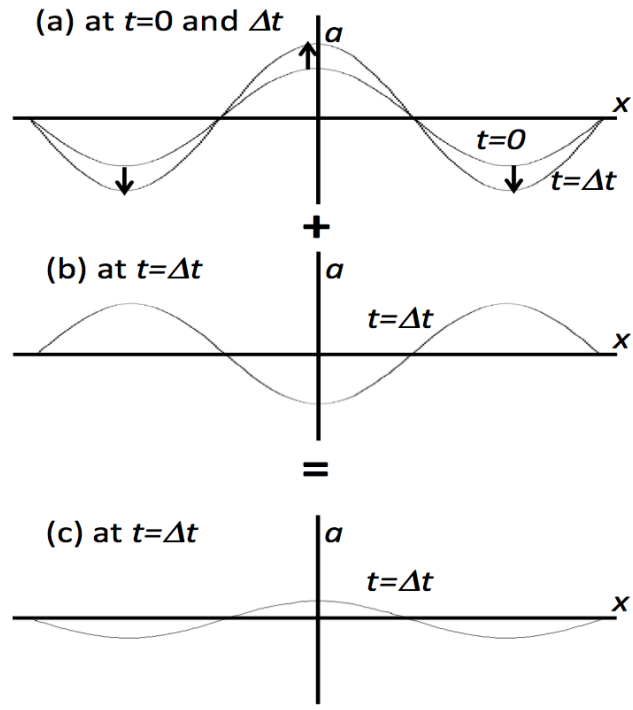

Fig. 1 An example concept of feedback control. (a)At $t=0$ a perturbation is imposed. The initial perturbation may grow at instability onset. (b) After $\Delta t$, if the feedback control works on the system, another perturbation, which has an inverse phase with the detected amplitude at $t=0$, is actively imposed, so that (c) the actual perturbation amplitude is mitigated very well after the superposition of the initial and additional perturbations.

## 2. Dynamic stabilization of plasma instability under strong driving force oscillation

In Refs. [1-3] one dynamic stabilization mechanism was proposed to stabilize the R-T instability based on the strong oscillation of acceleration, which was realized, for example, by the picket fence pulse train or the laser intensity modulation in laser inertial fusion [4]. In this mechanism, the total acceleration strongly oscillates, so that the additional oscillating force is added to create a new stable window in the system. Originally this dynamic stabilization mechanism was proposed by P. L. Kapitza [23], and it was applied to the stabilization of an inverted pendulum. The inverted pendulum is an unstable system, and on the system a strongly and rapidly oscillating acceleration is applied in Ref. [23], and then the inverted pendulum system has a stable window. In this case, the equation for the unstable system is modified, and has another force term coming from the oscillating acceleration. In this mechanism, the growth rate is modified by the strongly oscillating acceleration.

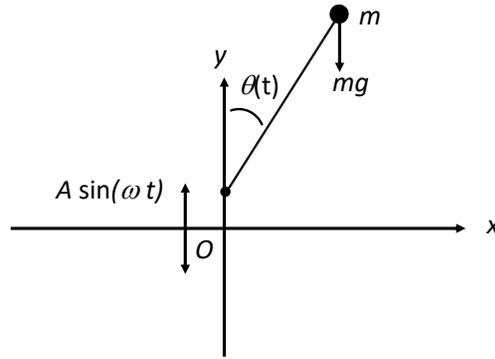

Fig. 2 Kapitza's pendulum, which can be stabilized by applying an additional strong and rapid acceleration of $A \sin \omega t$.

When the inverted pendulum shown in Fig. 2 is subjected by a strongly oscillating acceleration of $A \sin \omega t$, we obtain the following Mathieu-type equation [24] for $\theta(t)$:

$$\frac{d^2\theta(t)}{dt^2} = \frac{g}{l}\theta(t) - A\omega^2 \theta(t) \sin \omega t \qquad (1)$$

Here $l$ is the length of the pendulum. When $A=0$, the inverted pendulum becomes unstable. However, the second term of the righthand side is added to the system, and stable windows appear in the inverted pendulum system [23, 24]. In Eq. (1) the stability condition is described as $A - 0.5 < 2g/(l\omega^2) < A^2$ .[24] The stability condition shows that the additional acceleration oscillation at the second term of the righthand side of Eq. (1) should be very fast, and the amplitude of $A$ must satisfy the stability condition.

This dynamic stabilization mechanism works on, for example, the inverted pendulum in Fig. 1. However, it would be difficult to apply this mechanism to our tall buildings, bridges or large structures in our society.

In laser inertial fusion, this dynamic stabilization mechanism was proposed and applied to stabilize the R-T instability based on the strong oscillation of acceleration [3, 4], which was realized by the picket fence pulse train or the laser intensity modulation in laser inertial fusion [4]. In this mechanism, the total acceleration strongly oscillates, so that the additional oscillating force is added to create a new stable window in the fuel pellet implosion in laser inertial fusion. In inertial fusion, the spherical fuel pellet should be compressed to a high density, for example, a thousand times of the solid density [8-10]. The fusion fuel is imploded spherically by a large inward acceleration. The typical implosion acceleration is about $10^{13}$m/s$^2$, and lasts for about ns~10ns. During the implosion time, the driver input energy, introduced by the laser-pulse train series, would induce the strong implosion acceleration oscillation, which contributes to stabilize the RTI during the fuel pellet implosion [3, 4, 25, 26].

In Ref. [27], this dynamic stabilization mechanism is applied to the two-stream instability stabilization, in which the classical two-stream instability driven by a constant relative drift velocity is modified by the additional oscillation on the relative velocity. The time-dependent drift velocity opens a new stable window in the two-stream instability.

## 3. Dynamic stabilization of plasma instability under a phase control

In plasmas the perturbation phase and amplitude cannot be measured dynamically. However, by using a wobbling beam or an oscillating beam or a rotating beam or so[18, 19], the initial perturbation is actively imposed so that the initial perturbation phase and amplitude are defined actively. In this case, the amplitude and phase of the unstable perturbation cannot be detected, but can be defined by the input driver beam wobbling at least in the linear phase. In plasmas it would be difficult to realize a perfect feedback control, but a part of it can be adapted to the instability mitigation in plasmas. Actually, heavy ion beam accelerators would provide a controlled wobbling or oscillating beam with a high frequency [18-20, 28]. An intense electron beam axis can be also wobbled in its controlled way, and thus provides defined phase and amplitude of perturbations.

If the energy driver beam wobbles uniformly in time, the imposed perturbation for a physical quantity of $F$ at $t = \tau$ may be written as

$$F = \delta F e^{i\Omega\tau} e^{\gamma(t-\tau)+i\vec{k}\cdot\vec{x}}. \tag{2}$$

Here $\delta F$ is the amplitude, $\Omega$ the wobbling or oscillation frequency, and $\Omega\tau$ the phase shift of superimposed perturbations. At each time $t = \tau$, the wobbler provides a new perturbation with the controlled phase shifted and amplitude defined by the driving wobbler itself. After the superposition of the perturbations, the overall perturbation is described as

$$\int_0^t d\tau \, \delta F e^{i\Omega\tau} e^{\gamma(t-\tau)+i\vec{k}\cdot\vec{x}} \propto \frac{\gamma+i\Omega}{\gamma^2+\Omega^2} \delta F e^{\gamma t} e^{i\vec{k}\cdot\vec{x}}. \tag{3}$$

At each time of $t = \tau$ the driving wobbler provides a new perturbation with the shifted phase. Then each perturbation grows with the factor of $e^{\gamma t}$. At $t > \tau$ the superimposed overall perturbation growth is modified as shown above. When $\Omega \gg \gamma$, the perturbation amplitude is reduced by the factor of $\gamma/\Omega$, compared with the pure instability growth ($\Omega = 0$) based on the energy deposition nonuniformity [12, 21, 22, 29].

Figure 3 shows the superimposed perturbations decomposed, and at each time the phase-defined perturbation is imposed actively by the driving wobbler. The perturbations are superimposed at the time $t$. The wobbling trajectory is under control by for example a beam accelerator or so, and the superimposed perturbation phase and amplitude are controlled so that the overall perturbation growth is also controlled.

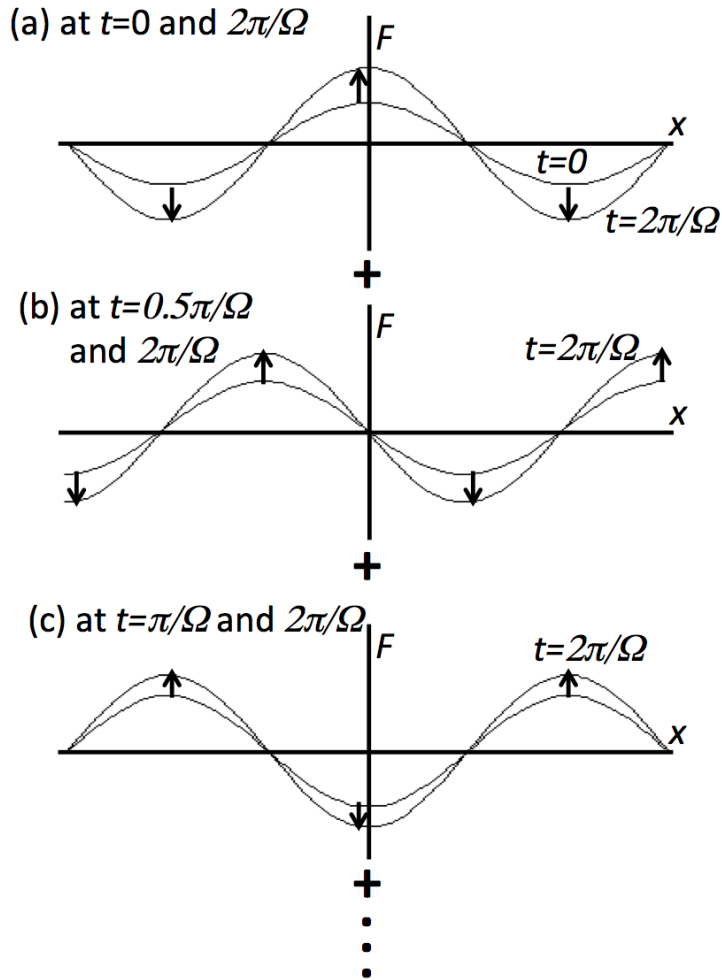

Fig. 3 Superposition of perturbations defined by the wobbling driver beam. At each time the wobbler provides a perturbation, whose amplitude and phase are defined by the wobbler itself. If the system is unstable, each perturbation is a source of instability. At a certain time the overall perturbation is the superposition of the growing perturbations. The superimposed perturbation growth is mitigated by the beam wobbling motion.

From the analytical expression for the physical quantity $F$ in Eq. (3), the mechanism proposed in this paper does not work, when $\Omega \ll \gamma$. Only modes, fulfilling the condition of $\Omega \geq \gamma$, can experience the instability mitigation through a wobbling process. For RTI, the growth rate $\gamma$ tends to become larger for a short wavelength. If $\Omega \ll \gamma$, the modes cannot be mitigated. In addition, if there are other sources of perturbations in the physical system and if the perturbation phase and amplitude are not controlled, this dynamic mitigation mechanism also does not work. For example, if the sphericity of an inertial fusion fuel target is degraded, the dynamic mitigation mechanism does not work. In this sense the dynamic mitigation mechanism is not almighty. Especially for a uniform compression of an inertial fusion fuel all the instability stabilization and mitigation mechanisms would contribute to release the fusion energy.

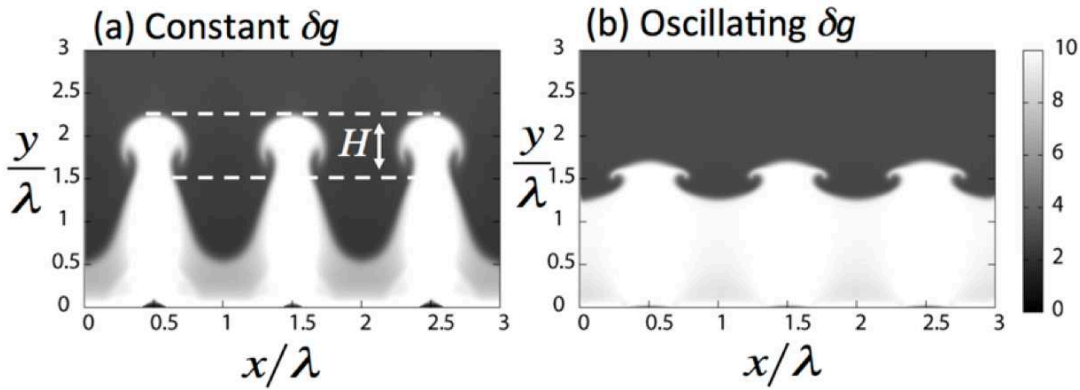

Fig. 4 Example simulation results for the Rayleigh-Taylor instability (RTI) mitigation. $\delta g$ is 10% of the acceleration $g_0$ and oscillates with the frequency of $\Omega=\gamma$. As shown above and in Eq. (2), the dynamic instability mitigation mechanism works well to mitigate the instability growth.

Figure 4 shows an example simulation for RTI, which has one mode. In this example, two stratified fluids are superimposed under an acceleration of $g = g_0 + \delta g$. In this example, two stratified fluids are superimposed under an acceleration of $g = g_0 + \delta g$. The density jump ratio between the two fluids is 10/3. In this specific case the wobbling frequency $\Omega$ is $\gamma$, the amplitude of $\delta g$ is $0.1 g_0$, and the results shown in Figs. 4 are those at $t = 5/\gamma$. In Fig. 4(a) $\delta g$ is constant and drives the RTI as usual, and in Fig. 4(b) the phase of $\delta g$ oscillates with the frequency of $\Omega$ as stated above for the dynamic instability stabilization in this section. The RTI growth mitigation ratio is 72.9% in Fig. 4. The growth mitigation ratio is defined by ($H_0$ - $H_{mitigate}$)/$H_0$×100%. Here $H$ is defined as shown in Fig. 4(a), $H_0$ shows the deviation amplitude of the two-fluid interface in the case in Fig. 4(a) without the oscillation ($\Omega = 0$), and $H_{mitigate}$ presents the deviation for the other cases with the oscillation ($\Omega \neq 0$). The example simulation results support well the effect of the dynamic mitigation mechanism. The example simulation results also support well the effect of the dynamic mitigation mechanism. Other multi modes RTI analyses are found in Ref. [11].

In order to check the robustness of the dynamic instability mitigation mechanism [29], here we study the effects of the change in the phase, the amplitude and the wavelength of the wobbling perturbation $\delta F$, that is, $\delta g$ in Fig. 4 on the dynamic instability mitigation.

When the perturbation amplitude $\delta F = \delta F(t)$ depends on time or oscillates slightly in time, the dynamic mitigation mechanism is examined first. We consider $\delta F(t) = \delta F_0 \left(1 + \Delta e^{i\Omega' t}\right)$ in Eq. (1). Here $\Delta \ll 1$. In this case, Eq. (3) is modified as follows:

$$\int_0^t d\tau \, \delta F e^{i\Omega\tau} e^{\gamma(t-\tau)+i\vec{k}\cdot\vec{x}} \propto \left\{\frac{\gamma+i\Omega}{\gamma^2+\Omega^2} + \Delta \frac{\gamma+i(\Omega+\Omega')}{\gamma^2+(\Omega+\Omega')^2}\right\} \delta F_0 e^{\gamma t} e^{i\vec{k}\cdot\vec{x}} \quad (4)$$

When $\Delta \ll 1$ in Eq. (4), just a minor effect appears on the dynamic mitigation of the instability.

We also performed the fluid simulations. In the simulations $\delta g(t) = \delta g(1 - \Delta sin\Omega' t)$. The RTI is simulated again based on the same parameter values shown in Fig. 4 except the perturbation amplitude oscillation $\delta F(t)$. In the simulations we employ $\Omega' = 3\Omega$, $\Omega$ and $\Omega/3$ in Eq. (4). For $\Delta$=0.1 and 0.3, and for $\Omega' = 3\Omega$, $\Omega$ and $\Omega/3$, the RTI growth reduction ratio is 54.9~73.2% at $t = 5/\gamma$. Figure 5 shows the results for $\Delta$=0.3. The results by the fluid simulations and Eq. (4) demonstrate that the perturbation amplitude oscillation $\delta F = \delta F(t)$ is uninfluential as long as $\Delta \ll 1$.

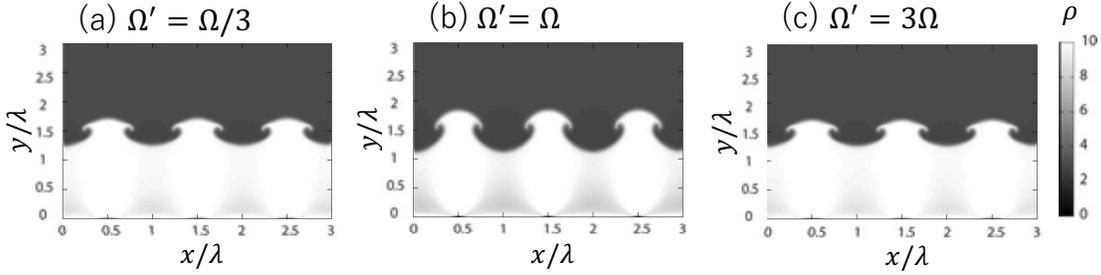

Fig. 5 Fluid simulation results for the RTI mitigation for the time-dependent $\delta g(t) = \delta g - \Delta \sin \Omega' t$ at $t = 5/\gamma$. In the simulations $\Delta = 0.3$, and (a) $\Omega' = \Omega/3$, (b) $\Omega' = \Omega$ and (c) $\Omega' = 3\Omega$. The dynamic mitigation mechanism is robust against the time change of the perturbation amplitude $\delta g(t)$.

When the oscillation frequency $\Omega$ of the perturbation $\delta F$ depends on time ($\Omega = \Omega(t)$), the time-dependent frequency means that $\Omega(t)$ would consist of multiple frequencies: $e^{i\Omega t} = \sum_i \Delta_i e^{i\Omega_i t}$. In this case Eq. (3) becomes

$$\int_0^t d\tau \, \delta F e^{i\Omega \tau} e^{\gamma(t-\tau)+i\vec{k}\cdot\vec{x}} \propto \sum_i \Delta_i \frac{\gamma + i\Omega_i}{\gamma^2 + \Omega_i^2} \delta F e^{\gamma t} e^{i\vec{k}\cdot\vec{x}}. \tag{5}$$

The result in Eq. (5) shows that the highest frequency of $\Omega_i$ contributes to the instability mitigation. In a real system the highest frequency would be the original wobbling frequency $\Omega$ or so, and the largest amplitude of $\Delta_i$ is also that for the original wobbling mode. So when the frequency change is slow, the original wobbler frequency of $\Omega$ contributes to the mitigation.

The fluid simulations are also done for the RTI with $\Omega(t) = \Omega\left(1 + \Delta \sin \Omega' t\right)$ together with the same parameter values employed in Fig. 4. In this case $\Delta=0.1$ and 0.3, and $\Omega' = 3\Omega$, $\Omega$ and $\Omega/3$. The growth reduction ratio was 66.9~74.0% at $t = 5/\gamma$. Figure 6 presents the simulation results for $\Delta=0.3$. The little oscillation of the imposed perturbation oscillation frequency $\Omega(t)$ has a minor effect on the dynamic instability mitigation.

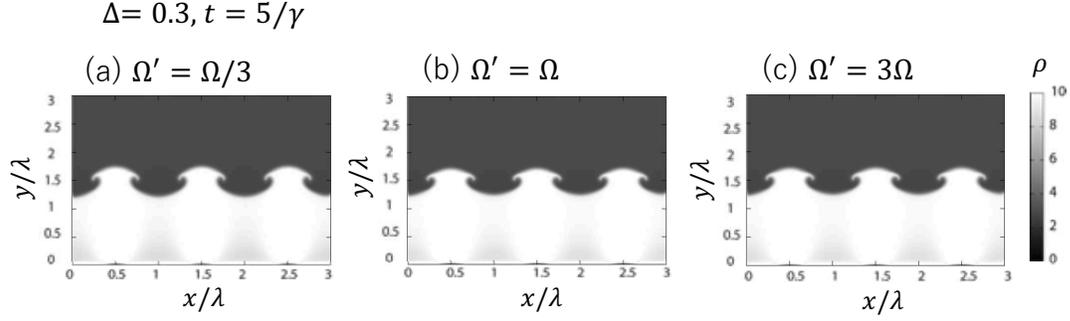

Fig. 6 Fluid simulation results for the RTI mitigation for the time-dependent wobbling frequency $\Omega(t) = \Omega_0(1 + \Delta \sin \Omega' t)$ at $t = 5/\gamma$. In the simulations $\Delta = 0.3$, and (a) $\Omega'_0 = \Omega/3$, (b) $\Omega'_0 = \Omega$ and (c) $\Omega'_0 = 3\Omega$. The dynamic mitigation mechanism is also robust against the time change of the perturbation frequency $\Omega(t)$.

When the wobbling wavelength $\lambda = 2\pi/k$ depends on time, one can expect as follows in a real system: $k(t) = k_0 + \Delta k e^{i\Omega'_k t}$ and $k_0 \gg \Delta k$. In this case the wobbling wavelength changes slightly in time, and Eq. (3) becomes as follows:

$$\int_0^t d\tau\, \delta F e^{i\Omega \tau} e^{\gamma(t-\tau)+ik\cdot x} \propto \delta F e^{\gamma t + ik_0\cdot x} \int_0^t d\tau\, e^{(i\Omega-\gamma)\tau} \sum_{m=-\infty}^{\infty} i^m J_m(\Delta k \cdot x) e^{im\Omega'_k \tau}$$

$$\propto \sum_{m=-\infty}^{\infty} i^m J_m(\Delta k \cdot x) \int_0^t d\tau\, e^{i(\Omega+m\Omega'_k)\tau - \gamma\tau} \propto \sum_{m=-\infty}^{\infty} i^m J_m(\Delta k \cdot x) \frac{\gamma + i(\Omega + m\Omega'_k)}{\gamma^2 + (\Omega + m\Omega'_k)^2} \quad (6)$$

Here $J_m$ is the Bessel function of the first kind. The result in Eq. (6) demonstrates that the instability growth reduction effect is not degraded by the small change in the wobbling wavelength. In actual situations the mode $m = 0$ contributes mostly to the instability mitigation, and in this case the original reduction effect shown in Eq. (3) is recovered.

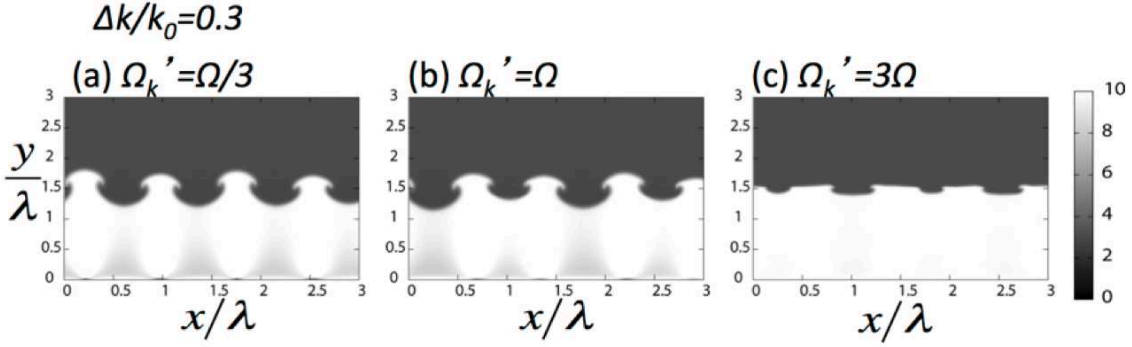

Fig.7 Fluid simulation results for the RTI mitigation for the time-dependent wobbling wavelength $k(t) = k_0 + \Delta k e^{i\Omega'_k t}$ at $t = 5/\gamma$. In the simulations $\Delta k/k_0 = 0.3$, and (a) $\Omega'_k = \Omega/3$, (b) $\Omega'_k = \Omega$ and (c) $\Omega'_k = 3\Omega$. The dynamic mitigation mechanism is also robust against the time change of the perturbation wavelength $k(t)$.

The fluid simulations are also performed for this case $k(t) = k_0 + \Delta k e^{i\Omega'_k t}$. Figure 7 shows the example simulation results for $\Delta k/k_0 = 0.3$ and $\Omega'_k = 3\Omega$, $\Omega$ and $\Omega/3$. Figure 7(a) shows the RTI growth reduction ratio of 61.3% for $\Omega'_k = \Omega/3$, Fig. 7(b) shows 68.0% for $\Omega'_k = \Omega$, and Fig. 7(c) shows 93.3% for $\Omega'_k = 3\Omega$ at $t = 5/\gamma$. For a realistic situation $\Omega'_k \sim \Omega$, where $\Omega$ is the wobbling or modulation frequency.

All the results shown above demonstrate that the dynamic instability mitigation mechanism proposed is rather robust against the changes in the amplitude, the phase and the wavelength of the wobbling or modulating perturbation of $\delta F$ in general or $\delta g$ in RTI.

Another possible example is the filamentation instability [13-16, 22] as shown in Fig. 8 schematically. In this example an electron beam is injected into a plasma, and the electron beam has a density or current modulation in transverse. The modulation is the source of the perturbation defined actively by the electron beam itself, and so the perturbation phase is defined. From the initial perturbation the filamentation instability grows with its growth rate. In this filamentation instability a magnetic field perturbation is induced by the electron beam modulation, the electron trajectories are bent and then the electron beam perturbation is further enhanced so that the magnetic field is also enhanced. If the electron beam axis oscillates transversally, the perturbations, which could have different phase, are successively imposed in the system and the dynamic mitigation mechanism works.

It is assumed that an electron beam moving in the $x$ direction with $v_{be}$ has a small density perturbation in the transverse direction ($y$). The perturbed electron beam is injected into a plasma as shown in Fig. 9. The current density perturbation induces the filamentation instability [13-16, 22], in which the perturbation of the transverse magnetic field in the $z$ direction grows, the electron trajectories are bent by the magnetic field, and consequently the current perturbation is enhanced.

The growth rate of the filamentation instability is expressed by $\gamma_F \cong \beta\sqrt{\alpha/\gamma_b}\omega_{pe}$, where $\beta = v_{be}/c$, $\alpha = n_{be}/n_p$, $n_{be}$ is the electron beam number density, $n_p$ the number density of the background plasma electrons, $\gamma_b = 1/\sqrt{1-\beta^2}$, and $\omega_{pe}$ the plasma frequency of the background plasma electrons.

Figure 9 shows the dynamic stabilization mechanism for the filamentation instability schematically. The input electron beam is injected into a plasma, and the electron beam has a current modulation in the *y* direction. The electron beam current modulation defines actively the filamentation phase as shown in Fig. 9(a). After a short time of $\Delta t$, the filamentation instability grows. Then the electron beam oscillates in the *y* direction as shown in Fig. 9(b), and the electron beam modulation also moves in the *y* direction. The new perturbation with the shifted phase is applied, and the perturbations grow. The overall instability growth should be defined by the sum of all the perturbations at *t*, and the filamentation instability is dynamically stabilized as shown in Fig. 1(c).

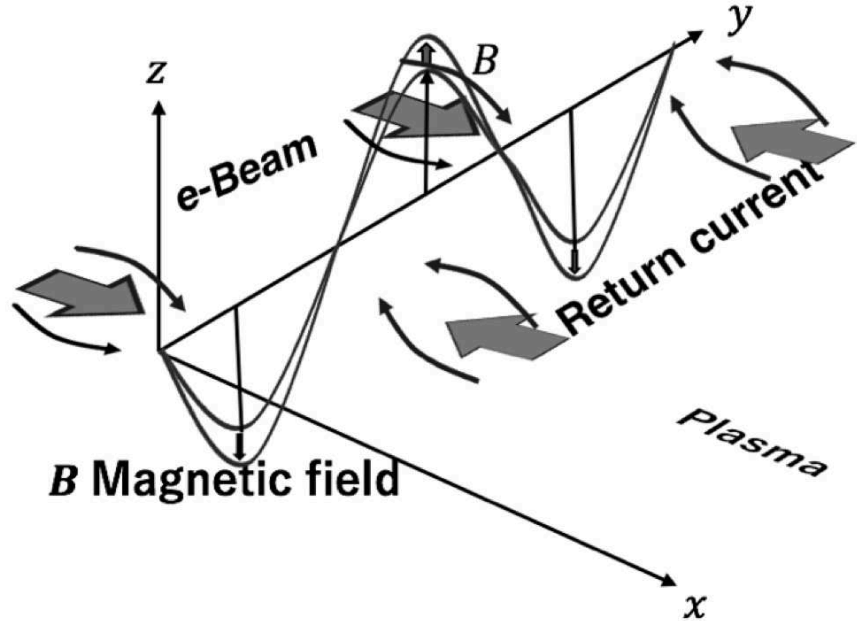

Fig. 8 Filamentation instability. In this case an electron beam has a density perturbation in transverse, and is injected into a plasma. In the plasma return current is induced to compensate the electron beam current. The perturbed electron beam itself defines the filamentation instability phase, and the e-beam axis oscillates in the y direction in this example case. Therefore, the filamentation instability is mitigated by the dynamic stabilization mechanism shown in Fig. 1 and in the second section.

In order to verify the filamentation instability stabilization, we perform 2-dimensional particle-in-cell simulations. As an example case, we use the following parameter values: $\alpha = n_{be}/n_p = 1/9$, $\beta = v_{be}/c = 0.9$, $v_{pe}/c = -0.1$, the temperatures of the beam electrons, the background electrons and the background ions are 100eV. In our simulations, $n_p = 1.00 \times 10^{-3} \times 4\pi^2\epsilon_0 m_e c^2/(\lambda e)^2$, the time is normalized by $1/\omega_{pe}$ and the scale length is normalized by $\lambda$.

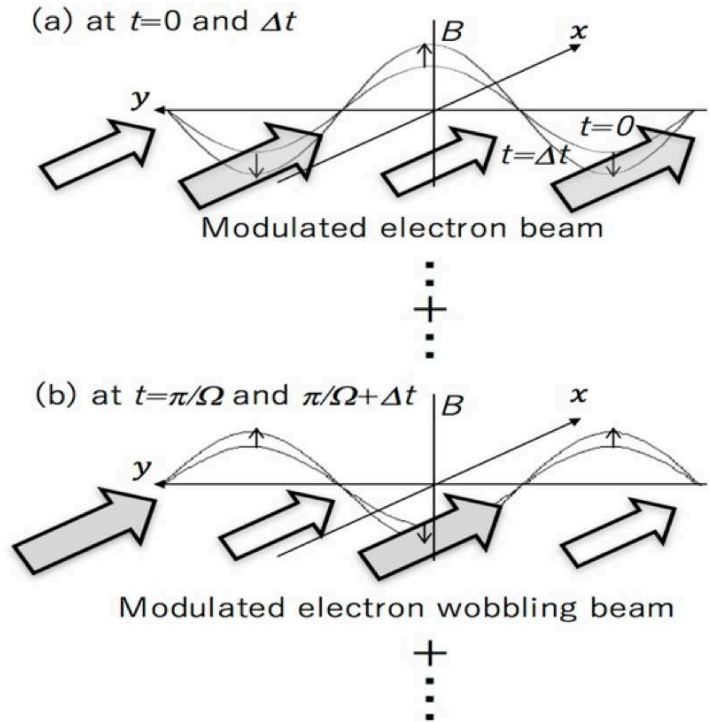

Fig. 9 Dynamic stabilization mechanism for the filamentation instability. (a) A modulated electron beam is imposed to induce the filamentation instability. The electron beam axis is wobbled or oscillates transversally with its frequency of $\Omega$. (b) At a later time its phase-shifted perturbation is additionally imposed by the electron beam itself. The overall perturbation is the superimposition of all the perturbations, and the filamentation instability is dynamically stabilized.

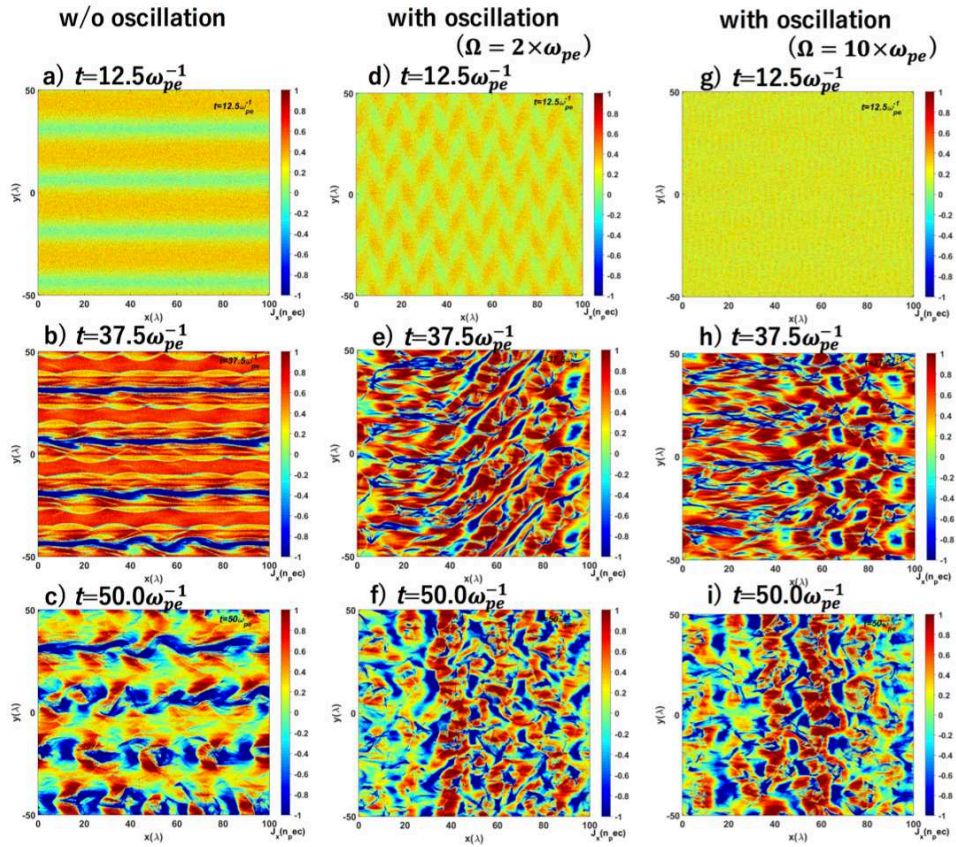

Fig. 10 Filamentation instability simulation results without and with the electron beam oscillation. The current density $J_x$ is shown at each time step. When the electron beam axis oscillates in the $y$ direction ( see Figs. d)-f) and g)-i) ), the filamentation instability growth is clearly mitigated.

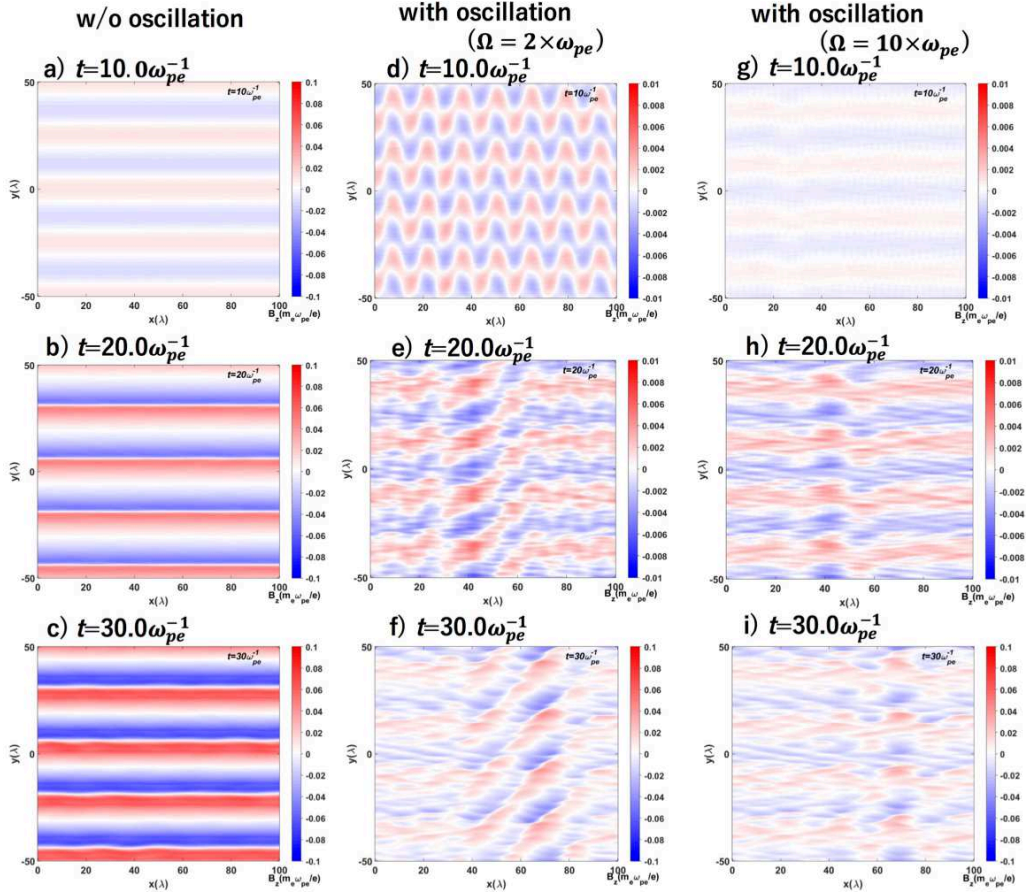

Fig. 11 Magnetic field $B_z$ for the filamentation instability without and with the electron beam oscillation.

Figures 10-12 show the simulation results for the filamentation instabilities with and without the electron beam oscillation. The electron beam perturbation is imposed in the beam density, and the amplitude is 10%. The oscillation amplitude is $5\lambda$ in the $y$ direction in these specific cases. The electron beam oscillation frequency $\Omega$ is $2\omega_{pe}$, $10\omega_{pe}$ and $20\omega_{pe}$ ($> \gamma_F$). Figure 10 presents the current density for the cases without and with the electron beam oscillation in the $y$ direction. Figures 11 show the magnetic field $B_z$ distribution. The stabilization effect of the filamentation instability is clearly demonstrated in Figs. 11. Figure 12 shows the magnetic field energy history. The dynamic stabilization ratio is introduced by $R_r = \left(1 - (U_{Bz}/U_{Bz0})\right) \times 100$, where $U_{Bz}$ shows the magnetic field energy. $U_{Bz}$ is normalized by the magnetic field energy $U_{Bz0}$ obtained without the electron beam oscillation. At $t = 35\omega_{pe}^{-1}$, the stabilization ratio of $R_r =$58.6% in the case of $\Omega=2\omega_{pe}$. When the electron beam transverse oscillation frequency

$\Omega$ in the $y$ direction becomes larger than or comparable to $\gamma_F$, the dynamic stabilization effect is remarkable. In addition, we have also performed a 3D simulation for the filamentation instability stabilization, under the same parameter values shown in Figs. 10-12, for

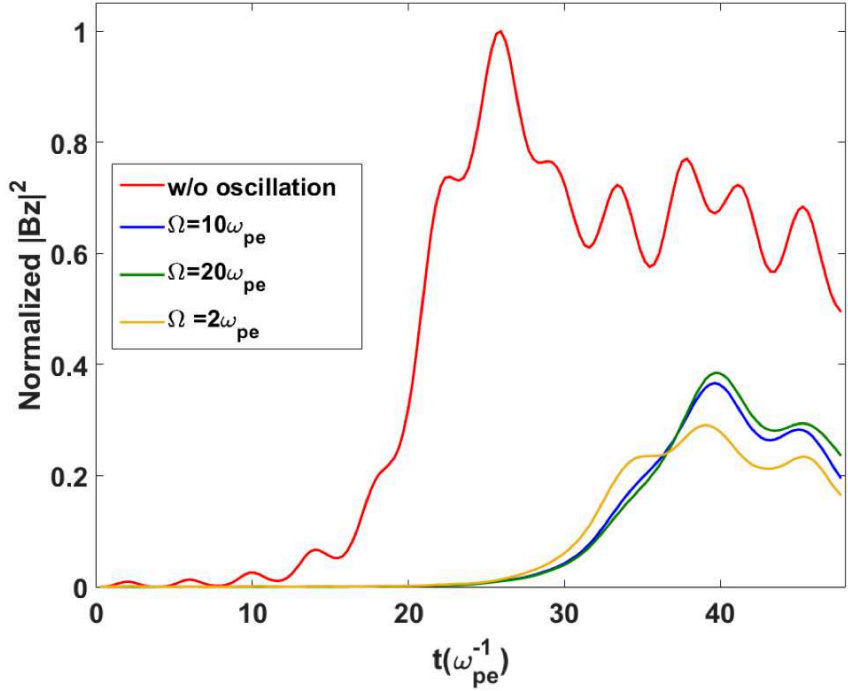

Fig. 12 Histories of the normalized magnetic field energy $U_{Bz} \propto |B_z|^2$. When the electron beam transverse oscillation frequency $\Omega$ in $y$ becomes larger than or comparable to $\gamma_F$, the dynamic stabilization effect is remarkable.

$\Omega=2\omega_{pe}$ with the circular rotation of the electron beam axis trajectory with the amplitude of $2\lambda$. The results are shown in Figs. 13(a) and (b). The 3D results also support the theoretical and 2D simulation results, and present that the initial clear filament structure is mitigated by the electron axis oscillation as shown in Fig. 13(b). The results shown in Figs. 10-13 demonstrate that the dynamic stabilization mechanism works well to stabilize the filamentation instability.

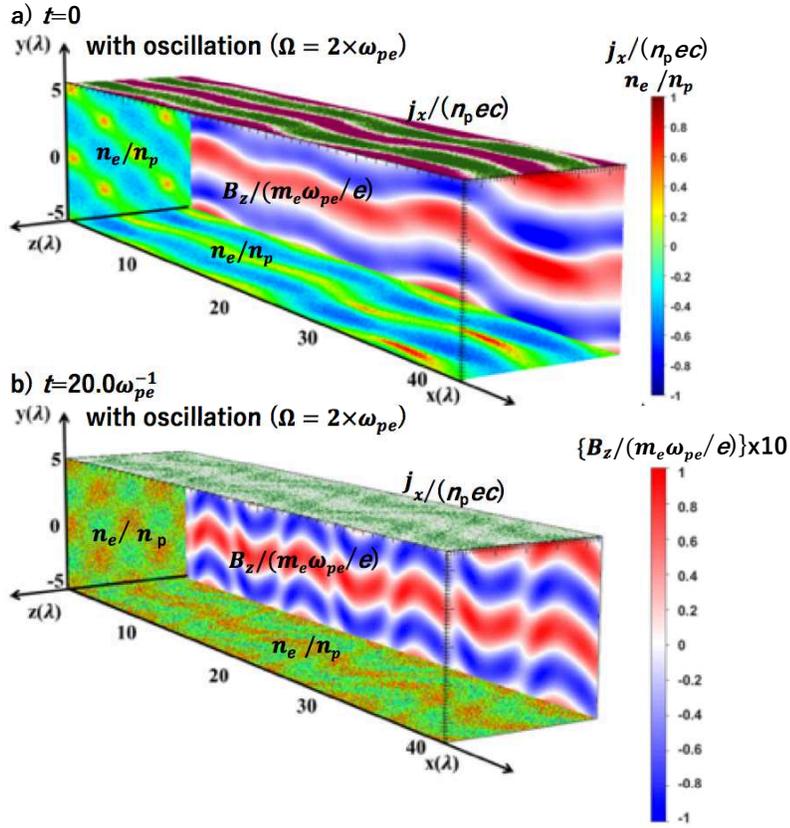

Fig. 13 3D simulation results on dynamic stabilization mechanism for the filamentation instability. (a) The initial setup, and (b)$n_e$, $j_x$ and $B_z$ are presented at t=20/$\omega_{pe}$. The initial clear filaments are gradually mitigated by the electron beam oscillation. The filamentation instability is dynamically stabilized.

## 4. Discussions and summary

In this paper we have discussed on the dynamic stabilization in plasmas. The dynamic stabilizations [1-4], based on the "Kapitza's pendulum" [23], introduce a new strong oscillating force into the basic equation, and then the governing equation is modified by the additional term to create a new stable window in the system. Therefore, the growth rate is modified, and the stable window appears in the system. The dynamic stabilization mechanism has been applied to the inverted pendulum [23], to a fuel target implosion in laser inertial fusion [4], and also to the stabilization of the two-stream instability [27]. Another dynamic stabilization mechanism, which is also based on the strong forced field but is different from the "Kapitza's pendulum", was also proposed and applied to a new field in a dissipative dynamic system to find a stable region in the system [30-32]. On the other hand, the dynamic stabilization mechanism based on the phase control was proposed and applied to the stabilization of plasma instabilities including the RTI, the filamentation instability, and also the fuel target implosion in heavy ion inertial fusion (HIF) [20-22, 29]. Originally the dynamic stabilization mechanism comes from the imperfect feedback control, which is widely used to stabilize tall building, structures, etc. in our society. In the perfect feedback control, the displacement and its phase are measured, and the additional perturbation is added to stabilize the systems. In plasmas we cannot measure the perturbation phase and amplitude. As we discussed in this paper, we can actively apply the perturbations. Then before moving to the system disruption or before developing to the non-linear phase, the additional perturbations, which should have reverse phase, are applied actively, so that the superimposed total amplitude would be mitigated.

**Acknowledgements:** This work was partly supported by MEXT, JSPS Kakenhi 15K05359, ILE/Osaka University, CORE / Utsunomiya University, and Japan-U.S. Fusion Research Collaboration Program conducted by MEXT, Japan. This work was also partially supported by the project HiFi (CZ.02.1.01/0.0/0.0/15_003/0000449) and project ELI: Extreme Light Infrastructure (CZ.02.1.01/0.0/0.0/15_008/0000162) from European Regional Development. Computational resources were partially provided by the ECLIPSE cluster of ELI Beamlines. The Authors would like to appreciate to X. F. Li, H. Katoh, J. Limpouch, O. Klimo, D. Margarone, Q. Yu, Q. Kong, S. Weber, S. Bulanov, and A. Andreev for their fruitful discussions on this subject.